Summary

Hungary's convergence to the developed western economies have been much slower than initially expected. Applying the FOI model, this study investigates whether there were any changes in the convergence process during the second decade of the 21$^{st}$ century. It is found that the future (influencing the long-term competitiveness of the economy), and inside (determining the current well-being of the country) potential of the Hungarian economy did not improve at all compared to the 34 countries that were OECD members in 2010. Hungary's position is in fact really bad, it is ranked 33rd in both areas. The country does somewhat better in the outside potential (characterising the world market position), prompting that Hungary follows a growth model that is focused on external resources. This feature is not new, however: the same development model patterns were detected in 2010, too.

Introduction

Hungary joined the OECD in 1996, and the EU in 2004. In the late 1980ies, when the transition process was started, the general expectation was that the economic convergence of Hungary will be relatively quick. By the mid 1990ies, however, the high hopes were much reduced; Balázs in his 1997 paper remarked: "the results achieved lag behind the expectations of the candidate countries" (Balázs 1997, p. 954). Economic convergence has been a highly debated topic ever since (e.g. Csaba 2011; Lengyel and Kotosz 2018; Győrffy 2022). I will only mention a few general figures to illustrate the difficulties one faces when discusses the topic: in 1996 the Hungarian PPP GDP/capita was 41% of the Austrian, 53% of the Irish, and 134% of the Polish value; by 2022 it moved to 63% (Austria), 33% (Ireland), and 98% (Poland) (IMF WEO 2022). Austria has been in the crosshairs of highly ranked Hungarian government officials for the past decade anyway. György Matolcsy, the governor of the Hungarian National Bank famously claimed that Hungary may well catch up to Austria within a couple of decades during the release of his book in 2016 (H.J. 2016), and he repeated this claim many times in the following years; in 2022 newly nominated cabinet ministers also mentioned the general aim of converging to Austria's/EU's level by 2030 (Sztojcsev 2022).

Not only the mere question of convergence is debated, but there is also disagreement on the development path Hungary has taken since its accession to the OECD/EU. The governor of the Hungarian National Bank persistently argues that Hungary has taken a revolutionary new path since 2010 (Matolcsy 2021), which is in line with the suggestion of a "Hungarian" model of development (Víg 2014). Győrffy (2022) on the other hand concludes that the 2010s have not brought significantly new policies and models for Hungary, the country is stuck in the middle income trap, and still pursues the cost-based growth model. The communique that suggests 144 policy reforms released by the Hungarian National Bank in 2022 seems to agree with Győrffy: although it states that the 2010s have been a golden age for the economy, it also remarks that Hungary has been lagging behind in productivity, digitisation, and smart, sustainable solutions (MNB 2022).

This study seeks answers two these two questions: 1) Is there evidence for a convergence in the Hungarian economy over the 2000-2020 period, and 2) What are the characteristics of the Hungarian development path? The individual contribution of my study is that it uses a comparative method by evaluating the performance of Hungary relative to the performance of other OECD members and it detects the individual traits of the development path by considering the future, outside and inside pillars of development using the FOI model.

The rest of this study is divided into four parts. First it gives a short literature review on Hungary's development over the past two decades; then it introduces the FOI model and the data sources used. In part three the model calculations are presented, finally, in part four I offer a discussion of the results.

Short literature review

Most researchers agree that in GDP/capita terms Hungary has been converging to the EU average over the 2000-2020 period, with a major slowdown in the years before, during, and after the great financial crisis. There seems to be two debated issues related to Hungary's development, however. Many authors claim that the growth of the Hungarian economy obscures the fact that what was achieved is just simply not good enough. Furthermore, a very concerning feature of the Hungarian growth is that the growth model chosen is not sustainable, and so it does not guarantee the long-term convergence of the country. Authors often having a vested interest in the Hungarian government on the other hand like to emphasize the extraordinary, and exceptional nature of the Hungarian growth following the financial crisis.

Baksa and Kónya (2021) evaluate the convergence patterns of the Visegrad 4 countries (V4), and Slovenia over the period of 1996-2019 using a stochastic, neoclassical growth model. They find a general convergence throughout most of those 25 years and remark that the EU funds have played a considerable part in the growth and investment patterns. Hungary was characterised with a positive investment climate until 2007, but the period also led to a high level of indebtedness which caused a major reversion and meant that Hungary had the highest interest rates until 2017. Policy changes in the early 2010s lead to an increase in the labour supply and a more flexible labour market.

Tóth (2019) concludes that although the GDP/capita of Hungary has been getting closer to the EU mean since the early 2000s (rising from 60% to 70%), the number of years required to reach the mean is still extremely high (35-50 years).

Regional level analysis can be one of the ways to explain why the picture painted about the Hungarian growth trends is so bleak. Wołkonowski (2019), over the 2004-2015 period, investigates the beta-convergence of the countries that joined the EU in 2004, or later on different levels: country, NUTS1, NUTS2, and NUTS3. The strongest evidence of beta-convergence is found on the country level, however Wołkonowski concludes that the lower the level of the unit investigated, the weaker the beta-convergence level gets. On the NUTS3 level he even finds that that there is divergence in case of the Hungarian regions.

Szakálné Kanó and Lengyel (2021) also focus on the NUTS3 regions of the V4 countries for the period of 2000-2016. They cluster the NUTS3 regions into 5 clubs by testing the convergence of their relative transition path. While Budapest, and Győr-Moson-Sopron are clustered into club 2 that shows a strong convergence, all the other Hungarian NUTS3 regions (basically: counties) went into club 4 (11 counties), and 5 (6 counties), the two worst groups characterised by poorly educated labour force, high unemployment, and decreasing population. One county (Nógrád) was such an outlier that it could not even be clustered into any of the five clubs.

Another possible reason why researchers are pessimistic about the convergence of the Hungarian economy is the broader picture, when indicators of the socioeconomic environment are also considered. Soreg and Bermudez-Gonzalez (2021) select three development pillars (Economic and business performance, Socio-political performance, and Human development and quality of life performance), and use 21 different indicators to measure the performance of Hungary, Croatia, Romania, and Bulgaria between 1996 and 2019. They find that all four countries show a relative high dependence on foreign direct investment (FDI), have a dual economic structure, which makes them very vulnerable to exogenous shocks. By comparing these countries to the other members of the V4 Soreg and Bermudez-Gonzalez conclude that Hungary has been converging to the Balkan-three, and has been falling behind the V4 countries.

Bokros (2021) who has been one of the masterminds behind the mid-1990s macroeconomic stabilisation reform, goes much further, and suggests that following the 1995 reform Hungary had been on a sustainable growth path, which deteriorated in the first decade of the 2000s, and starting from the 2010s the decade of decay set in. This, of course, is in stark contrast with the views of Matolcsy (2021), who believes that the 2010s has been the most successful decade of the Hungarian economy in the last 100 years.

Given that the two authors are positioned at the two opposing sides of the political spectrum, their disagreement does not come as a major surprise. The contradicting view on the Hungarian model comes in a period of policy convergence. As Bielik (2021) shows, the economic activity of the V4 countries is not significantly related to who governs the country, as he finds that the economic effect of governments characterised by different political ideologies on economic activity was neither substantial, nor meaningful.

One of the lessons of this review is that indicators signalling about the deeper layers of the Hungarian socioeconomic environment can offer a better understanding about the development path the country has taken. This study addresses this point by adopting the FOI model as the main method of analysis.

Methodology and data sources

In my study I calculate the position of Hungary, and the other OECD members along three main development pillars (future, outside, and inside pillars) in 2000, 2010, and 2020. The methodology of the FOI model was described in detail in Bartha and Gubik (2014), I only summarise the main points here.

1. The model is based on the idea that institutional factors determine the long term development path of a country. To identify the traits of the development path taken by Hungary, the FOI model is set up that measures the future, outside, and inside potential of the economy. The future potential considers the long-term competitiveness of the economy; the outside potential determines the current world market position of the economy; while the inside potential summarizes factors that are crucial for the current well-being of the community.
2. The three potentials are measured using the F, O, and I indices.
   a. Eleven variables are used to derive the value of the F index, namely: a measure of social responsibility/sustainability, labour market cooperation, flexibility of labour force, reliability of energy infrastructure, expenditure on education, aging of the society, share of renewable energy, life expectancy, ecological footprint, expenditure on research & development, efficiency of the education system.
   b. Five variables contribute to the O index: trade openness, country credit rating, financial sector stability, exchange rate stability, foreign language skills.
   c. The eight variables involved in the I index: efficiency of government intervention, quality of life, tax revenues, pension system stability, GDP/capita, entrepreneurial soundness, labour market flexibility, availability of skilled labour.
   
   The 2000, and 2010 FOI calculations were done almost a decade ago, and some of the variables used for the indices have changed since then. The 2020 FOI indices therefore are slightly different from the previous ones, but the discontinued variables were replaced by ones with similar content, so this should only cause a minor distortion in the output.
3. In 2022 there are 38 members in the OECD – an increase of 4 since 2010; since some of the variables required are not available for the years 2000, and 2010 in case of the four new members, I only included the 34 countries in the analysis that were already OECD members in 2010.
4. To calculate the three indices, all variable values are standardised to a 1-7 scale using the minmax method. There are a total of 24 variables (11 for F, 5 for O, and 8 for I), and ideally all variables should have 34 values for the 34 countries for the years 2000, 2010, and 2020. For every year and variable there is a best and a worst value; these best/worst values are used to calculate the standardised value of the variable using this simple formula: 6*(actual country value – worst value)/(best value – worst value)+1. This formula converts all figures to a 1-7 scale, where the country with the best value will have standardised value of 7, and the country with the worst value has a standardised value of 1.
5. The FOI indices are then calculated as the mean of the standardised values of the variables belonging to the F, O, and I pillar. The country that has the highest F index in 2020 (Iceland with 5.3) can then be declared as the country with the highest future potential for growth.
6. The obtained indices can be used for several purposes. In this study I use them to conduct a cluster analysis, and this helps to identify development paths among OECD members. The indices are also a relative measure of a country's performance (compared to the best/worst performing OECD member). The index score, as well as the country rank according to the score can be used to check the development level of a country.

The variables used for the calculations were obtained from the following sources:
1. OECD.Stat: https://stats.oecd.org/
2. WEF Global Competitiveness Report (Schwab 2019)

Results and discussion

Table 1 lists all the F, O, and I-index scores for the 34 OECD countries for 2000, 2010, and 2020. The table also shows the rank of the country among the 34 members according to the given index. The rank is probably an even better expression of a country's performance, as it not only dependent on the best, worst and own variable values, but it also reflects the changes taken place in all the other countries.

Table 1. The FOI index scores for 34 OECD members in the years 2000, 2010, and 2020

| Country | F-2020 | F-2010 | F-2000 | O-2020 | O-2010 | O-2000 | I-2020 | I-2010 | I-2000 |
|---|---|---|---|---|---|---|---|---|---|
| Australia | 3.8 (24.) | 4.6 (13.) | 4.5 (18.) | 5.3 (4.) | 5.3 (10.) | 4.6 (11.) | 4.6 (12.) | 4.4 (6.) | 4.3 (14.) |
| Austria | 4.4 (10.) | 5.1 (9.) | 5.3 (7.) | 5.1 (8.) | 5.4 (8.) | 4.2 (16.) | 3.9 (18.) | 4 (12.) | 4.7 (7.) |
| Belgium | 3.8 (22.) | 4.2 (17.) | 5.1 (11.) | 4.9 (14.) | 5.6 (5.) | 4.9 (7.) | 3.6 (22.) | 3.5 (21.) | 4.3 (16.) |
| Canada | 4 (17.) | 4.2 (18.) | 4.9 (15.) | 4.9 (11.) | 5.4 (7.) | 5 (4.) | 4.6 (11.) | 4.5 (2.) | 4.7 (8.) |
| Chile | 3.6 (27.) | 3.8 (21.) | 3.9 (23.) | 3.9 (29.) | 5 (14.) | 4 (20.) | 3.8 (19.) | 4.1 (9.) | 2.9 (31.) |
| Czechia | 3.8 (25.) | 3.4 (27.) | 3.1 (31.) | 4.2 (25.) | 5 (15.) | 2.4 (33.) | 3.2 (25.) | 3.6 (20.) | 3.3 (27.) |
| Denmark | 4.9 (4.) | 5.3 (8.) | 5.2 (9.) | 5 (10.) | 5.8 (2.) | 4.4 (14.) | 4.7 (9.) | 4.3 (7.) | 4.8 (5.) |
| Estonia | 4.2 (16.) | 3.2 (30.) | 3.1 (30.) | 4.7 (16.) | 4.9 (16.) | 3.7 (22.) | 3.6 (21.) | 3.1 (25.) | 3.3 (26.) |
| Finland | 4.6 (7.) | 5.4 (7.) | 5.6 (5.) | 5.1 (9.) | 5.7 (3.) | 4.6 (12.) | 4.9 (6.) | 4 (13.) | 5.1 (2.) |
| France | 4.2 (15.) | 4.7 (12.) | 5 (13.) | 4.3 (22.) | 4.5 (21.) | 4 (19.) | 3.5 (23.) | 3 (27.) | 4.3 (15.) |
| Germany | 4.4 (11.) | 4.8 (11.) | 4.9 (14.) | 4.7 (17.) | 5.3 (11.) | 4.3 (15.) | 4.5 (15.) | 3.7 (18.) | 4.3 (13.) |
| Greece | 3.3 (30.) | 3.1 (31.) | 3 (32.) | 2.9 (34.) | 3.7 (32.) | 2.8 (31.) | 1.9 (34.) | 2.5 (34.) | 3.2 (29.) |
| Hungary | 3.1 (33.) | 3.2 (29.) | 3.4 (28.) | 4.4 (21.) | 4.6 (19.) | 3.2 (26.) | 2.6 (33.) | 2.5 (33.) | 3.4 (24.) |
| Iceland | 5.3 (1.) | 5.8 (3.) | 5.6 (2.) | 4.2 (24.) | 2.3 (34.) | 4.1 (17.) | 5 (4.) | 4.4 (5.) | 5.1 (3.) |
| Ireland | 4.3 (14.) | 4.2 (19.) | 4.1 (20.) | 4.6 (18.) | 4.2 (28.) | 4.7 (10.) | 5 (5.) | 3.9 (16.) | 4.5 (12.) |
| Israel | 4.5 (9.) | 3.6 (26.) | 4.2 (19.) | 4.6 (19.) | 4.9 (17.) | 4.1 (18.) | 4.1 (17.) | 4.1 (10.) | 4.3 (17.) |
| Italy | 3.5 (28.) | 3.7 (22.) | 3.9 (24.) | 3.5 (32.) | 3.8 (30.) | 3.2 (28.) | 2.7 (32.) | 2.7 (32.) | 3.6 (21.) |
| Japan | 4.7 (6.) | 5.5 (5.) | 5.6 (3.) | 3.7 (30.) | 3.7 (31.) | 3.5 (24.) | 4.1 (16.) | 4 (14.) | 3.5 (22.) |
| Korea | 4.3 (12.) | 4.5 (14.) | 4 (22.) | 4.3 (23.) | 4.3 (26.) | 3.5 (25.) | 3.8 (20.) | 3.3 (22.) | 3.3 (28.) |
| Luxembourg | 3.8 (23.) | 6.1 (1.) | 5.4 (6.) | 6.1 (1.) | 6.6 (1.) | 5.8 (1.) | 4.6 (13.) | 4.5 (4.) | 5.7 (1.) |
| Mexico | 3 (34.) | 2.6 (34.) | 3 (33.) | 4.1 (26.) | 4 (29.) | 3 (30.) | 3.3 (24.) | 2.9 (30.) | 2.4 (34.) |
| Netherlands | 4.3 (13.) | 4.9 (10.) | 5.1 (10.) | 5.3 (6.) | 5.5 (6.) | 5 (3.) | 5.3 (2.) | 3.8 (17.) | 4.6 (9.) |
| New Zealand | 4.5 (8.) | 4.4 (15.) | 4.7 (17.) | 5.1 (7.) | 4.5 (20.) | 4.5 (13.) | 4.8 (8.) | 4 (15.) | 4.1 (18.) |
| Norway | 4.7 (5.) | 5.5 (4.) | 5.2 (8.) | 4.9 (13.) | 5.7 (4.) | 5 (5.) | 4.9 (7.) | 4.1 (11.) | 4.6 (10.) |
| Poland | 3.7 (26.) | 3.1 (32.) | 3.2 (29.) | 4 (28.) | 4.4 (22.) | 3.2 (29.) | 3.1 (29.) | 3.1 (26.) | 2.8 (32.) |
| Portugal | 3.9 (19.) | 3.7 (25.) | 3.6 (26.) | 3.7 (31.) | 4.3 (24.) | 3.9 (21.) | 3.1 (28.) | 2.9 (29.) | 3.4 (23.) |
| Slovakia | 3.4 (29.) | 3.3 (28.) | 3.6 (27.) | 4.8 (15.) | 4.8 (18.) | 2.6 (32.) | 2.9 (31.) | 3.3 (23.) | 3.1 (30.) |
| Slovenia | 4 (18.) | 3.7 (23.) | 4.1 (21.) | 4.5 (20.) | 5.1 (13.) | 3.2 (27.) | 3.2 (26.) | 2.7 (31.) | 3.3 (25.) |
| Spain | 3.2 (31.) | 3.7 (24.) | 3.7 (25.) | 4 (27.) | 4.2 (27.) | 3.7 (23.) | 3.1 (27.) | 3 (28.) | 4 (20.) |
| Sweden | 4.9 (3.) | 5.5 (6.) | 5.6 (4.) | 4.9 (12.) | 5.2 (12.) | 4.8 (9.) | 4.6 (14.) | 4.1 (8.) | 4.7 (6.) |
| Switzerland | 5.2 (2.) | 5.9 (2.) | 5.9 (1.) | 5.4 (3.) | 5.4 (9.) | 4.8 (8.) | 5.6 (1.) | 4.9 (1.) | 4.9 (4.) |
| *Turkey* | 3.1 (32.) | 3 (33.) | 2.9 (34.) | 3.2 (33.) | 3.6 (33.) | 1.9 (34.) | 3.1 (30.) | 3.1 (24.) | 2.6 (33.) |
| *UK* | 3.8 (21.) | 4.3 (16.) | 4.8 (16.) | 5.3 (5.) | 4.3 (23.) | 5 (6.) | 4.7 (10.) | 3.6 (19.) | 4.1 (19.) |
| *USA* | 3.9 (20.) | 4.1 (20.) | 5 (12.) | 5.4 (2.) | 4.3 (25.) | 5 (2.) | 5.3 (3.) | 4.5 (3.) | 4.5 (11.) |

Source: own calculations

As the three potentials of the FOI model were introduced to capture the different aspects of development, one would expect that the rank of the countries differed according to the three indices. This is true in case of some countries (e.g. Iceland is among the best in the future, and inside pillar and towards the back in the outside pillar; Canada on the other hand is highly ranked according to O, & I, but is close to the middle in tis F rank), but there are countries with very similar ranking numbers as well (e.g. Switzerland, Luxembourg, or Norway are very highly ranked, while Greece, and Turkey are very lowly ranked in all categories). Hungary has quite low rankings, although the country seems to perform somewhat better in its outside potential.

Table 2 helps us to get a clearer picture about Hungary's relative position. The inside potential (the level of well-being of the population) had dropped significantly during the 2000s, and it has not recovered since then. The future potential (measuring the long term competitiveness of the economy) was rather low to begin with and the ranking of Hungary has been continuously dropping. The outside potential (the world market position) of Hungary is stronger than the other two pillars, it places the country in the midfield.

Table 2. Hungary's rank according to the FOI indices

| Year | F rank | O rank | I rank |
|---|---|---|---|
| 2020 | 33$^{rd}$ | 21$^{st}$ | 33$^{rd}$ |
| 2010 | 29$^{th}$ | 19$^{th}$ | 33$^{rd}$ |
| 2000 | 28$^{th}$ | 26$^{th}$ | 24$^{th}$ |

Source: own calculations

The rankings shown by Table 2 seem to back those researchers who concluded that Hungary's convergence has stalled. The measures of well-being put Hungary among the worst within the OECD, and the chance of a recovery is small as the future potential of the country is also very poor. The relatively higher ranking in outside potential suggests that Hungary continues to rely heavily on external resources in its development model.

In order to identify the special traits of the Hungarian development path within the OECD, a hierarchical cluster analysis was conducted using the 2020 FOI indices as variables. Between groups linkage was used as the cluster method with Squared Euclidean distance measure of intervals. Since the OECD had 38 members in 2020, all 38 countries were included in the clusters. Table 3 shows the countries that go into the same cluster as Hungary as we change the number of clusters from 3 to 11. The countries closest to Hungary based on their FOI index scores are Slovakia, Spain, Mexico, and Poland.

Table 3. Members of Hungary's cluster and their relative proximity to Hungary

| Number of clusters | Cluster members |
|---|---|
| 3-8 | Belgium (1.72), Chile (1.95), Czechia (0.9), Estonia (2.27), France (2), Italy (0.97), Korea (2.85), Latvia (0.84), Lithuania (1.32), Mexico (0.5), Poland (0.79), Portugal (1.53), Slovakia (0.33), Slovenia (1.19), Spain (0.44) |
| 9-10 | Chile (1.95), Czechia (0.9), Italy (0.97), Latvia (0.84), Lithuania (1.32), Mexico (0.5), Poland (0.79), Portugal (1.53), Slovakia (0.33), Spain (0.44) |
| 11 | Slovakia (0.33) |

Source: own calculations

Although it would be better to obtain more homogeneous groups, the 3-cluster classification seems to be the best option when using the hierarchical cluster method. Compared to the 3-cluster solution, Hungary is only moved to a smaller group if the number of clusters is increased to 9 but this is a very unbalanced classification where 2 clusters only have 2 members, and there are 4 countries that form individual clusters (Iceland, Japan, Luxembourg, and Switzerland).

Table 4 features the FOI index means for the three clusters. Hungary is in cluster 2. At first glance, the clustering does not seem to reveal major insights about the development paths of countries: cluster 1 has the highest index scores in all three potentials, cluster 2 has lower ones, while the lowest means belong to cluster 3. The one feature of Table 4 worth highlighting is the fact that cluster 2 has an above the average outside potential, meaning that it is not only Hungary but a whole group of countries that base their development on external resources. This is what Győrffy (2022) calls cost-based growth model but this finding is also in line with those studies that found that Hungary relies heavily on FDI and EU funds (Baksa and Kónya 2021; Soreg and Bermudez-Gonzalez 2021).

Table 4. FOI index means for the 3-cluster solution

| Cluster | F-index | O-index | I-index |
|---|---|---|---|
| 1 | 4,5 | 5,0 | 4,7 |
| 2 | 3,7 | 4,2 | 3,3 |
| 3 | 3,2 | 3,1 | 2,5 |

Source: own calculations

To better distribute the outliers among the dominant clusters, I tried an artificial clustering method, the result of which is shown in Table 5. The indices have a value between 1 and 7 meaning that 4 is the middle value. Every country can have a High (index is larger than 4) or a Low (index is lower than 4) index value in all three potentials, which means that a half-scale method can also be used to classify the countries into different groups. As we have three indices, the possible numbers of clusters is 8 but in 2020 37 of 38 OECD members went into 5 of these artificial, half-scale clusters.

Table 5. Clusters of OECD countries according to the half-scale method

| Cluster | Members |
|---|---|
| High F, O, I (FOI) | Denmark, Finland, Germany, Iceland, Ireland, Israel, Netherlands, New Zealand, Norway, Sweden, Switzerland |
| High F, O, low I (FOi) | Austria, Estonia, France, Korea |
| High F, I, low O (FoI) | Japan |
| Low F, high O, I (fOI) | Australia, Luxembourg, United Kingdom, United States |
| High F, low O, I (Foi) | - |
| Low F, I, high O (fOi) | Belgium , Czechia, Hungary, Latvia, Lithuania, Mexico, Poland, Slovakia, Slovenia, Spain |
| Low F, O, high I (foI) | - |
| Low F, O, I (foi) | Chile, Colombia, Costa Rica, Greece, Italy, Portugal, Turkey |

Source: own calculations

Hungary is in the fOi cluster, characterised by high outside and low future, and inside potential. Again, this does not come as a surprise, as similar features were found during the hierarchical clustering. Table 5 is still useful because it adds two extra information about development models. One is that the outside-focused development model is the second most common even among OECD members. The other added value of the half-scale clustering method is that it allows us to compare Hungary's current position to the one calculated for 2010 (Bartha and S. Gubik 2013).

Ten years ago we found that Hungary was also part of the fOi cluster, which was the second most populous group within the OECD back then, too (Bartha and S. Gubik 2013, p. 448). This suggests that there were no fundamental changes in the growth model of Hungary, so the 2010s did not bring major changes compared to the previous two decades. While Hungary's path seems to be unchanged, some countries did manage to make advances: Estonia, and especially Israel have moved to more sustainable development models; Mexico joined the fOi group from the low-everything cluster. There are examples for drop down as well: Belgium joined the Hungarian group from a more prestigious group, while Chile dropped to the cluster that previously included Mexico.

Conclusion

The two questions this study sought answer to were: 1) Is there evidence for a convergence in the Hungarian economy over the 2000-2020 period, and 2) What are the characteristics of the Hungarian development path? The analysis was conducted using the FOI model, and the results revealed that Hungary's relative position to the 34 OECD members did not improve significantly over the 2010-2020 period and, if anything, it got worse. The index measuring the future potential of the Hungarian economy ranked the country 33rd among the 34 countries (a drop of 4 places compared to 2010, and 5 places compared to 2000), and 33 was Hungary's rank according to the inside potential as well (just as in 2010; in 2000 it ranked 24th). The factors that measure the current level of well-being and influence the future sustainability of growth put Hungary towards the back end of the OECD members. The conclusion is that there are no signs of convergence when we compare Hungary's performance to the club of most developed countries.

Hungary does somewhat better in the outside potential (ranked 21st). The outside potential measures the world market position of an economy; countries with a high outside potential rely heavily on international markets and resources. But when the external focus is not paired with a high inside or future potential, the economy develops a dual structure, preventing the country from achieving long term convergence (Bartha and S. Gubik 2013). The outside-focused growth model therefore is considered to be fragile, leading to the so called middle income trap (Győrffy 2022). The external focus of the Hungarian development is not a novel thing; Hungary showed similar patterns in 2010, and even before that. The second conclusion therefore is that there are no major changes in Hungary's development model; what happened in the 2010s is not fundamentally different from what went on in the 1990s, and the 2000s.